\documentclass[11pt]{article}
\usepackage{epsfig,dsfont}
\usepackage{geometry}
\usepackage[normalem]{ulem}
\usepackage{color}

\usepackage{cite}

\begin{document}  
\sffamily

\vspace*{1mm}

\begin{center}

{\LARGE
Exploring the worldline formulation of the Potts model} \\
\vskip2mm
\vskip10mm
Christof Gattringer, Daniel G\"oschl, Pascal T\"orek
 
\vskip8mm
Universit\"at Graz, Institut f\"ur Physik, Universit\"atsplatz 5, 8010 Graz, Austria\footnote{Member of NAWI Graz.}
\end{center}
\vskip15mm

\begin{abstract}
We revisit the issue of worldline formulations for the $q$-state Potts model and discuss a worldline 
representation in arbitrary dimensions which also allows for magnetic terms. For vanishing magnetic field we 
implement a Hodge decomposition for resolving the constraints with dual variables, which in two dimensions 
implies self-duality as a simple corollary. We present exploratory 2-d Monte Carlo simulations in terms of the 
worldlines, based on worm algorithms. We study both, vanishing and non-zero magnetic field, and explore $q$ 
between $q = 2$ and $q = 30$, i.e., Potts models with continuous, as well as strong first order transitions.
\end{abstract}

\vskip15mm

\section{Introductory remarks}
The presentation of the Prokofev-Svistunov worm algorithm in \cite{worm} was a milestone for the developement of
worldline Monte Carlo simulations in statistical mechanics and lattice field theory. Not only did 
worldline\footnote{We point out that in the context of gauge theories the objects that correspond to worldlines
are actually discretized worldsheets described by occupation numbers placed on plaquettes. For notational 
convenience we continue to use ''worldlines'' but remark that often very similar techniques  
(see, e.g., \cite{Mercado:2013ola,Mercado:2013yta}) lead from worldlines for spins systems 
or lattice field theories to the corresponding ''worldsheet'' representations of gauge theories with the 
same symmetry group (which is of course a local symmetry in the gauge theory context).} 
representations and suitable algorithms considerably increase the numerical efficiency of Monte Carlo 
simulations, but also helped to solve some complex action problems, in particular those emerging in lattice 
field theories at finite chemical potential (for some reviews see, e.g., \cite{review1,review2,review3,review4}). 

It is important to note that there is no unique 
general approach to worldline techniques and suitable corresponding algorithms, but different types of systems 
have to be treated individually.  Over the years new tools and techniques have been developed for worldline 
representations and now also examples of worldline representations for some non-Abelian symmetries are known,
which pose a considerably more challenging task for a worldline approach. 

In this paper we explore worldline techniques for yet another   
system with a non-Abelian symmetry, the $q$-state Potts model \cite{potts} with the permutation group  
of $q$ elements as its symmetry group (see \cite{pottsreview} for a review). In the context of lattice field 
theories mainly the 3-state Potts model, which is equivalent to the $\mathds{Z}_3$ spin model, has been analyzed 
\cite{flux1} --\cite{Delgado:2011yp}. $\mathds{Z}_3$ is the center group of SU(3), the gauge group of QCD, such 
that the 3-state Potts model serves as a simple model theory for studying aspects of the strong interaction.

We here use a worldline representation for the Potts model at arbitrary $q$ that also allows for the inclusion of
magnetic terms, and work out the representation in arbitrary dimensions. The worldlines are 
represented by link-based fluxes $j_{x,\mu} \in \{ 0, 1, \, ... \, q-1\}$ that obey constraints which enforce flux conservation modulo $q$ at each site of the lattice. For vanishing field the worldlines of flux thus consist of closed loops with 
flux conservation modulo $q$. The magnetic field gives rise to additional sinks and sources for the worldlines,
such that in this case also open worldlines with magnetic terms at their endpoints are possible. 

For the case of vanishing magnetic field we implement a suitable form of the Hodge decomposition and 
show that the constraints can be obeyed by switching to dual variables
which are plaquettes of flux and defect lines of flux winding around the compactified directions of the lattice. For the special 
case of two dimensions the self-duality of the model at all $q$ follows directly from this dual 
representation of the worldlines
and one can re-derive the critical couplings $\beta_c = \ln(1+ \sqrt{q})$ \cite{betacrit} as a simple corollary.   

We finally also present a first exploratory assessment of the worldline representation for Monte Carlo simulations by 
discussing a suitable generalization of the worm algorithm. We analyze the worm algorithm in 2-d simulations where we
provide a determination of the dynamical critical exponent $z$ for the $q=2$ and $q=4$ cases which have second order 
phase transitions and for $q > 4$ address aspects of the worm algorithm for the emerging strong first order transitions.  

\section{Worldline representation for the Potts model}

In this section we will first derive the worldline form of the partition sum and discuss some observables in this 
representation which later will be the basis for simulations with the worm algorithm. For the case of vanishing magnetic field we 
will then introduce dual variables, i.e., plaquette variables and defect fluxes to resolve the constraints. This 
representation may be used for a local dual update that we will use for cross-checking the worm results. Finally 
we consider the 2-dimensional case where the dual representation can be used to establish full duality in the 
Kramers-Wannier sense \cite{Kramers:1941kn} that allows one to rederive the known results \cite{betacrit} for the critical 
values $\beta_{c}$ for all $q$.

\subsection{Derivation of the worldline representation}

We consider a $d$-dimensional hypercubic lattice with 
periodic boundary conditions. The sites of the lattice 
are denoted by $x$ and on each site we place a spin 
$s_x \in \{0, 1, \, ... \; q-1\}$ where $q \geq 2$ is a 
free integer-valued parameter. The partition sum of the 
$q$-state Potts model is given by 
\begin{equation}
Z \; = \; \sum_{\{ s \}} \; e^{ \; \beta \sum_{x}
\sum_{\mu =1}^d \delta(s_x,s_{x+\hat\mu}) \; + \; 
\eta \sum_{x} \delta(s_x,0)} \; = \;
\sum_{\{ s \}} \left[ \prod_{x,\mu} \; e^{ \; \beta 
\,\delta(s_x,s_{x+\hat\mu})} \right] 
\left[\prod_x \; e^{ \; \eta \, \delta(s_x,0)} \right],
\label{Zdef}
\end{equation}
where the parameter $\beta$ in front of the 
ferromagnetic nearest neighbor term is the 
inverse temperature and $\eta$ denotes an 
external magnetic field where the corresponding term
favors the spin orientation $s_x = 0$. $x$ is summed 
over all lattice sites, $\mu$ runs over all $d$ 
directions and $\hat \mu$ is the unit vector in
direction $\mu$.
$\sum_{\{s\}} = \prod_x \frac{1}{q}\sum_{s_x=0}^{q-1}$
is the sum over all spin configurations and we 
use $\delta(s,s^\prime) \equiv \delta_{s,s^\prime}$ 
to denote the Kronecker delta. In the second step of 
(\ref{Zdef}) we have rewritten the sums in the exponent 
into products of individual exponentials. 

Each of these exponentials can assume only two different
values, and we can write them as
\begin{equation}
e^{ \, \alpha \, \delta(s,s^\prime)} \; = \; 
[ e^{\, \alpha} - 1] \, \delta(s,s^\prime) + 1 \; = \;
[ e^{\, \alpha} - 1] \, \frac{1}{q} \sum_{j= 0}^{q-1}
e^{ \, i \, \frac{2\pi}{q} j \,(s-s^\prime)} + 1 \; = \;
\sum_{j=0}^{q-1} w_j^{\,\alpha} \; 
e^{\, i \, \frac{2\pi}{q}j \, (s-s^\prime)} \; ,
\label{exprep}
\end{equation}
where $\alpha$ is a real parameter. In the second step we have written the Kronecker 
delta as a sum over the $q$-th roots of unity. The 
weight factors $w_j^{\,\alpha}$ are given by 
\begin{equation}
w_0^{\,\alpha} \; = \; \frac{e^{\,\alpha} - 1}{q} + 1\; \; , 
\quad	w_1^{\,\alpha} \; = \; w_2^{\,\alpha} \; = \; ... \; 
w_{q-1}^{\,\alpha} \; = \; \frac{e^{\, \alpha} - 1}{q} \; .
\label{weightsw}
\end{equation}
We will use this representation of the exponentials both
for the nearest neighbor terms where we set 
$\alpha = \beta$, as well as for the magnetic terms where
$\alpha = \eta$ and $s^\prime = 0$. 

Using (\ref{exprep}) in the expression for the partition 
sum we obtain
\begin{eqnarray}
Z & = & \sum_{\{s\}} \left[ \prod_{x,\mu} 
\sum_{j_{x,\mu}=0}^{q-1} w^{\, \beta}_{j_{x,\nu}} 
e^{\,i \, \frac{2\pi}{q} \, j_{x,\mu} \,
(s_x - s_{x+\hat \mu})} \right] 
\left[ \prod_{x} 
\sum_{m_{x}=0}^{q-1} w^{\, \eta}_{m_{x}} 
e^{\,i \, \frac{2\pi}{q} \, m_{x} \, 
s_x } \right] 
\nonumber \\
& = & \sum_{\{ j \}} \sum_{\{ m \}} {\cal W}[j,m] 
\, \sum_{\{ s \}}  
\left[ \prod_{x,\mu} 
e^{\,i \, \frac{2\pi}{q} \, j_{x,\mu} \,
(s_x - s_{x+\hat \nu})} \right] 
\left[ \prod_{x} 
e^{\,i \, \frac{2\pi}{q} \, m_{x} \, 
s_x} \right] .
\label{step1}
\end{eqnarray}
For the representation of the Kronecker deltas in the 
nearest neighbor and magnetic terms we have introduced 
flux variables $j_{x,\mu} \in \{0,1, \, ... \; q-1\}$
and magnetic variables $m_x \in \{0,1, \, ... \; q-1\}$.
The sums over their configurations and the total weight 
are denoted as
\begin{equation}
\sum_{\{j\}} \; = \; \prod_{x,\mu} \; 
\sum_{j_{x,\mu} = 0}^{q-1} \; \; , \quad 
\sum_{\{m\}} \; = \; \prod_{x} \; 
\sum_{m_{x} = 0}^{q-1} \; \; , \quad
{\cal W}[j,m] \; = \; 
\left[ \prod_{x,\mu} w^{\, \beta}_{j_{x,\mu}}\right] \;
\left[ \prod_{x} w^{\, \eta}_{m_{x}}\right] \;.
\label{WLsum}
\end{equation}
The last step for obtaining the worldline 
representation of the Potts model is to sum over the
original spin variables in (\ref{step1}),
\begin{eqnarray}
&& 	\sum_{\{ s \}}  
\left[ \prod_{x,\mu} 
e^{\,i \, \frac{2\pi}{q} \, j_{x,\mu} \,
(s_x - s_{x+\hat \mu})} \right] 
\left[ \prod_{x} 
e^{\,i \, \frac{2\pi}{q} \, m_{x} \, 
s_x} \right] \; = \;
\, \sum_{\{ s \}}  
\prod_{x} 
e^{\,i \, \frac{2\pi}{q} \, s_x \, \left( 
\sum_\mu [j_{x,\mu} - j_{x-\hat\mu}] \, + \, 
m_{x} \right) }
 \\
&& \hspace*{30mm} = \; \prod_x \frac{1}{q} 
\sum_{s_x = 0}^{q-1} 
e^{\,i \, \frac{2\pi}{q} \, s_x \, \left( 
\sum_\mu [j_{x,\mu} - j_{x-\hat \mu}] \, + \, 
m_{x} \right) } \; = \; 
\prod_x \theta_q\left( 
\sum_\mu [j_{x,\mu} - j_{x-\hat \mu}] \, + \, 
m_{x} \right) ,
\nonumber 
\end{eqnarray}
where we have defined 
\begin{equation} 
\theta_q(n) \; = \;
\left\{ \begin{array}{ll}
1 & \mbox{if} \;\; n \, \mbox{mod} \, q \; = \; 0 \; ,
\\
0 & \mbox{else} \; . 	
\end{array} \right.
\end{equation}
Thus we find the worldline representation of the 
partition sum in the form
\begin{equation}
Z \; = \; \sum_{\{ j \}} \sum_{\{ m \}} {\cal W}[j,m]
\;\, \prod_x \, \theta_q \!\left( \, 
\vec \nabla \cdot \vec j_{x} \, + \, 
m_{x} \right) \; ,
\label{Z_WL}
\end{equation}
where we introduced the discretized divergence 
$\vec{\nabla} \cdot \vec j_{x}  = 
\sum_\mu [ j_{x,\mu} - j_{x-\hat{\mu},\mu}]$. 

\begin{figure}[t]
\begin{center}
\includegraphics[width=110mm,clip]{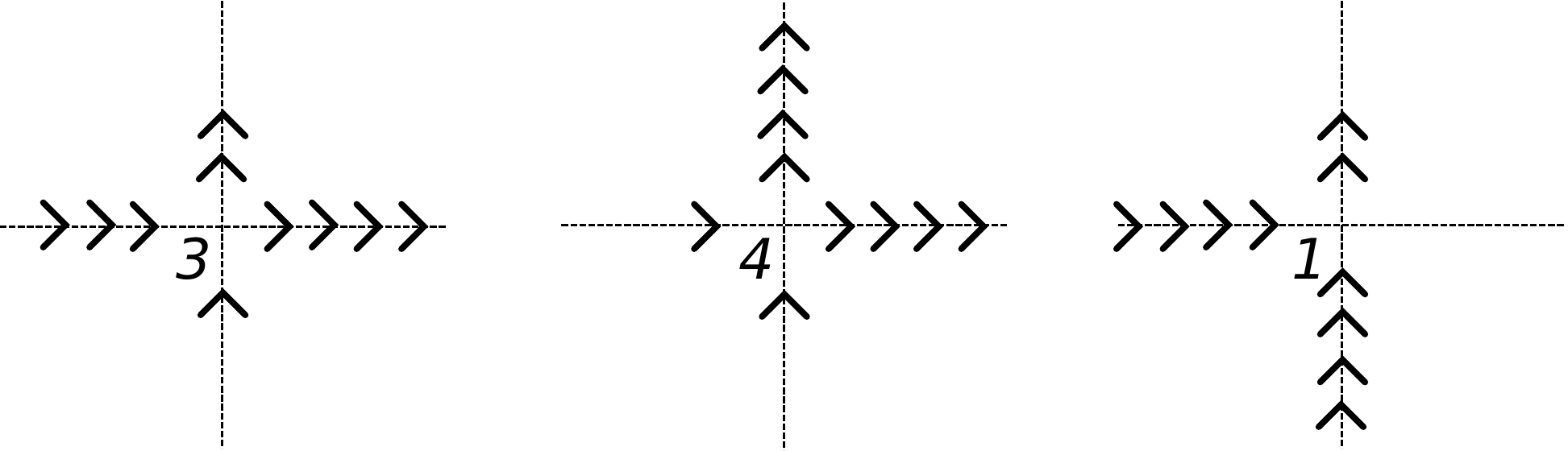}
\end{center}
\caption{Examples of admissible configurations of the
flux variables (shown as arrows on the links) and 
magnetic variables (numbers next to the central 
site) for the case of the $q = 5$ model. 
It is easy to check that all examples obey 
the constraint (\ref{constraints}).
\label{examples}}	
\end{figure}

The worldline form (\ref{Z_WL}) of the partition function is a sum over the configurations (\ref{WLsum})
of the flux variables $j_{x,\mu} \in \{0,1, \, ... \, q-1\}$ and the magnetic variables $m_x \in \{0,1, \, ... \, q-1\}$. 
Each configuration comes with a weight ${\cal W}[j,m]$ defined in (\ref{WLsum}). Admissible configurations 
of the flux and magnetic variables have to obey the constraints
\begin{equation}
\left( \vec \nabla \cdot \vec j_{x} \, + \, 
m_{x} \right) \, \mbox{mod} \; q \;\, = \,\; 0 
\; \; \; \; \forall \; x \; .	
\label{constraints}
\end{equation}
These constraints require that the sum of the divergence $\vec \nabla \cdot \vec j_x$ and 
the magnetic variable $m_x$ at each site $x$ has to vanish modulo $q$. This condition has the nice geometrical 
interpretation, that the flux $j_{x,\mu}$ is conserved modulo $q$ and that the magnetic variables
can act as sources and sinks modulo $q$. In Fig.~\ref{examples} we show for $q=5$ possible examples of flux and 
magnetic variables at a site $x$.  

The admissible configurations thus are worldlines of flux $j_{x,\mu}$ that is conserved 
modulo $q$ and the worldlines either form closed loops or open worldlines with magnetic terms $m_x$ 
at their endpoints which serve as sources or sinks (again modulo $q$).
Below we will discuss strategies for Monte Carlo updates of the flux configurations subject to the constraints 
(\ref{constraints}) based on the worm algorithm \cite{worm}. 

We conclude this section with providing the expressions for some bulk observables in terms of the flux and
magnetic variables. More specifically we consider the internal energy $E$ and the heat 
capacity $C$ which can be obtained as first and second derivatives of $\ln Z$ with respect to $\beta$, as 
well as the magnetization $M$ and the susceptibility $\chi$ which are the first and second derivatives 
with respect to $\eta$. These derivatives of $\ln Z$ can be evaluated directly in the worldline representation 
(\ref{Z_WL}) and one finds after a few steps of algebra
\begin{eqnarray}
E & = & -\frac{\partial}{\partial \beta} \, \ln Z \; = \; \left\langle - \sum_{x,\mu} 
\frac{w_{j_{x,\mu}}^{\, \beta\;\prime}}{w_{j_{x,\mu}}^{\, \beta}} 
\right \rangle \; , 
\label{observables} \\
C & = & \frac{1}{V} \frac{\partial}{\partial \beta} \, E \; = \; - \frac{1}{V} 
\left[ \left\langle \left(\sum_{x,\mu} 
\frac{w_{j_{x,\mu}}^{\, \beta \; \prime}}{w_{j_{x,\mu}}^{\, \beta}} \right)^{\!\!2} \; \right\rangle - 
\left\langle  \sum_{x,\mu} 
\frac{w_{j_{x,\mu}}^{\, \beta\;\prime}}{w_{j_{x,\mu}}^{\, \beta}}  \right\rangle^{\!\!2} + 
\left\langle \sum_{x,\mu} 
\frac{ w_{j_{x,\mu}}^{\,\beta\;\prime\prime} \, w_{j_{x,\mu}}^{\, \beta}  - 
w_{j_{x,\mu}}^{\, \beta\;\prime} \, w_{j_{x,\mu}}^{\, \beta\;\prime} }
{\left(w_{j_{x,\mu}}^{\, \beta}\right)^2}  \right\rangle \right] \; ,
\nonumber \\
M & = & \frac{\partial}{\partial \eta} \, \ln Z \; = \; \left\langle \sum_{x} 
\frac{w_{m_{x}}^{\,\eta\;\prime}}{w_{m_{x}}^{\, \eta}} 
\right \rangle \; ,
\nonumber \\
\chi & = & \frac{1}{V} \frac{\partial}{\partial \eta} \, M \; = \; \frac{1}{V} 
\left[ \left\langle \left(\sum_{x} 
\frac{w_{m_{x}}^{\, \eta \; \prime}}{w_{m_{x}}^{\, \eta}} \right)^{\!\!2} \; \right\rangle - 
\left\langle  \sum_{x} 
\frac{w_{m_{x}}^{\, \eta \; \prime}}{w_{m_{x}}^{\, \eta}}  \right\rangle^{\!\!2} + 
\left\langle \sum_{x} 
\frac{ w_{m_{x}}^{\,\eta\;\prime\prime} \, w_{m_{x}}^{\,\eta}  - 
w_{m_{x}}^{\,\eta\;\prime} \, w_{m_{x}}^{\,\eta\;\prime} }
{\left(w_{m_{x}}^{\, \eta}\right)^2}  \right\rangle \right] \; .
\nonumber
\end{eqnarray}
The expectation values on the right hand sides of (\ref{observables}) are expectation values in the worldline formulation.
$w_{j_{x,\mu}}^{\,\beta\;\prime}$ and $w_{j_{x,\mu}}^{\, \beta\;\prime\prime}$ denote the first 
and second derivatives of $w_{j_{x,\mu}}^{\, \beta}$ with respect to $\beta$, and $w_{m_{x}}^{\, \eta\;\prime}$ and  
$w_{m_{x}}^{\,\eta\;\prime\prime}$ the first and second derivatives of $w_{m_{x}}^{\,\eta}$ with respect to $\eta$. 

\subsection{Resolving the constraints with dual variables}

For the case of vanishing magnetic field, i.e., for $\eta = 0$, one may develop the worldline representation further 
by introducing {\it dual variables} such that the constraints are automatically fulfilled. For $\eta = 0$ the magnetic
sinks and sources are absent in the worldline representation and the admissible configurations of flux
correspond to closed loops, where at every site flux is conserved modulo $q$. The dual variables then consist of 
{\it plaquette variables} and {\it defect fluxes} that can be combined to build up all admissible configurations 
of the flux variables. 

\begin{figure}[t]
\begin{center}
\hspace*{-9mm}
\includegraphics[width=90mm,clip]{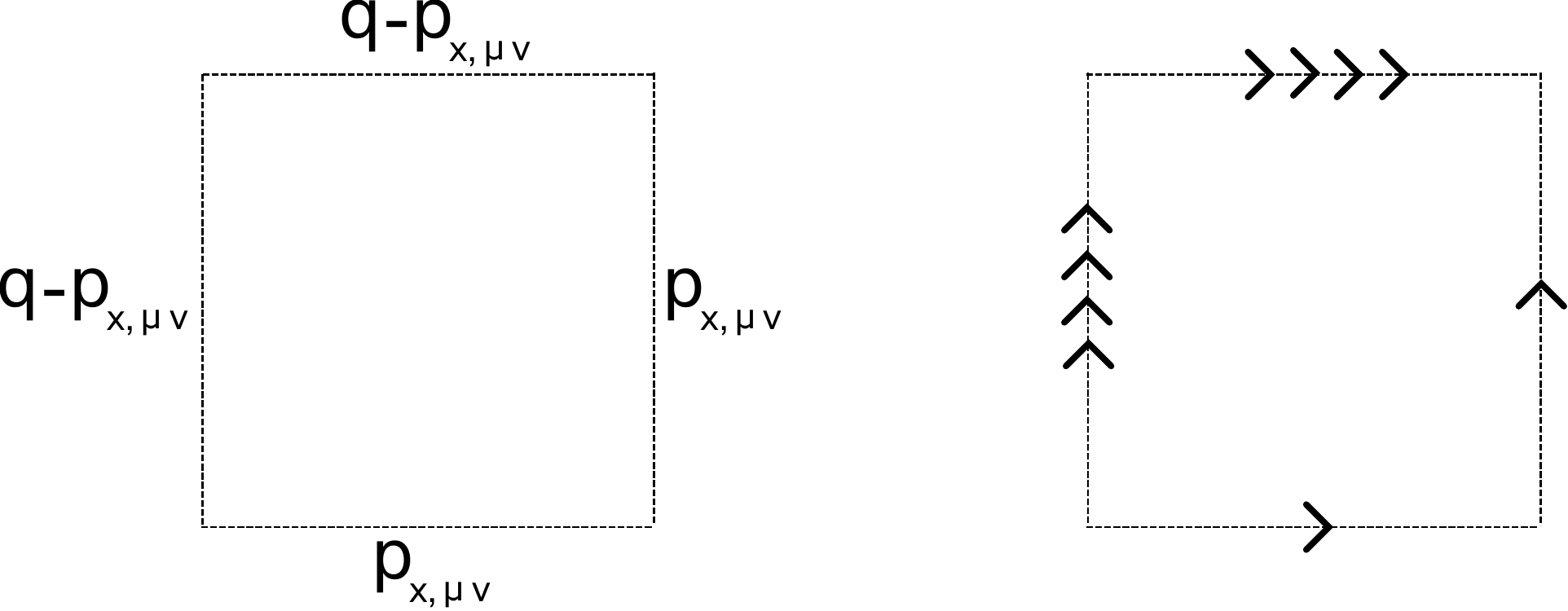}
\end{center}
\caption{Illustration of the dual plaquette variables: The lhs.~plot shows the general assignment of flux to the links of 
a plaquette with plaquette occupation number $p_{x,\mu \nu}$ for the case of the $q$-state model. The subtraction 
in the labeling of the links is understood modulo $q$ (compare (\ref{fluxplaq})). The rhs.~plot
gives an explicit example of the fluxes for $p_{x,\mu \nu} = 1$ in the $q=5$ model. 
\label{plaqvariables}}	
\end{figure}

The plaquette variables are integers 
$p_{x,\mu \nu} \in \{0,1, \, ... \; q-1\}$ placed on the plaquettes which we label as $(x,\mu, \nu)$ with 
$1 \leq \mu < \nu \leq d$.  A value $p_{x,\mu \nu}$ of a plaquette variable generates flux on the four links of the 
plaquette as follows\footnote{We remark that this assignment of fluxes to the links of the plaquettes is a convention and
other choices are possible that differ by multiples of $q$ on some of the links. The reason is  that the overall contribution 
of all plaquette variables 
to a link is always computed mod $q$ (see Eq.~(\ref{hodgedecomp})).}  
\begin{equation}
j_{x,\mu} \; \leftarrow \; p_{x,\mu \nu} \; , \; \;
j_{x+\hat{\mu},\nu} \; \leftarrow \; p_{x,\mu \nu} \; , \;\;
j_{x+\hat{\nu},\mu} \; \leftarrow \; (q-p_{x,\mu \nu})\; \mbox{mod} \; q \; , \;\;
j_{x,\nu} \; \leftarrow \; (q - p_{x,\mu \nu}) \; \mbox{mod} \; q \; .
\label{fluxplaq}
\end{equation} 
The configuration of flux that is generated by some $p_{x,\mu \nu} \neq 0$ obviously corresponds to the 
smallest possible closed loop with flux conserved modulo $q$. In Fig.~\ref{plaqvariables} we illustrate 
the loop using the general form of (\ref{fluxplaq}) in the lhs.\ plot and with the example $p_{x,\mu \nu}=1$ 
for $q=5$ in the rhs.\ plot.

At a given link $(x,\mu)$ the corresponding flux $j_{x,\mu}$ of course receives contributions from the plaquette 
variables of all plaquettes that contain $(x,\mu)$. To indicate that in (\ref{fluxplaq}) we only show the contribution of 
a single plaquette we used the notation with ''$\leftarrow$'' in (\ref{fluxplaq}). The full contribution of all plaquettes 
to a given link will then be summarized below. 

It is important to note that the flux generated by the plaquette variables does not generate all admissible flux 
configurations, since loops of conserved flux may also close around the periodic boundaries we use. This can be 
taken into account by introducing defect fluxes defined as
\begin{equation}
h_{x,\mu} \; = \; \sum_{\nu = 1}^d \delta_{\mu, \nu} \; f_\nu \; S_\nu(x) \; , \; \; 
\mbox{with} \; \; f_\nu \; \in \; \{ 0,1, \, ... \; q-1\}  \; ,
\end{equation}
and
\begin{equation}    
S_\nu(x) \; = \; \left\{ 
\begin{array}{l} 
1 \; \; \mbox{for} \; \;  x \; = \; \hat{\nu} \, j \,,  \; j \, = 0, 1, \, ... \; N_\nu-1 \; ,\\
0 \; \; \mbox{otherwise} \; .
\end{array}
\right. 
\end{equation} 
The $S_\nu(x)$ are the support functions for the coordinate axes in direction $\nu$ through the origin, and the 
parameters $f_\nu, \; \nu = 1,2, \, ... \; d$ determine the flux that is introduced on the corresponding loops
along the coordinate axes that close around the periodic boundary conditions. 

We now may represent all admissible configurations of the flux $j_{x,\mu}$ in the form
\begin{equation}
j_{x,\mu} \; = \; \left(- \!\sum_{\rho: \rho < \mu} [p_{x,\rho\mu} - p_{x-\hat \rho,\rho\mu}] \; + 
\sum_{\nu: \mu < \nu} [p_{x,\mu\nu} - p_{x-\hat \nu,\mu\nu}] \; + \; h_{x,\mu} \right) \mbox{mod} \; q\; .
\label{hodgedecomp}
\end{equation}
In a more abstract language, the representation (\ref{hodgedecomp}) is the Hodge decomposition \cite{hodge1,hodge2}
of all flux configurations with $(\vec \nabla \cdot \vec j_{x}) \, \mbox{mod} \, q = 0 \; \forall \, x$. 
In terms of the dual variables the partition sum now reads
\begin{equation}
Z \; = \; \sum_{\{p\}}  \sum_{\{f\}} {\cal W}[\,p,f] \; , \quad
\sum_{\{p\}} \; = \; \prod_{x} \prod_{\mu<\nu} \; \sum_{p_{x,\mu \nu}=0}^{q-1} 
\; , \quad
\sum_{\{f\}} \; = \; \prod_{\mu}  \; \sum_{f_{\mu}=0}^{q-1} \; , \quad
{\cal W}[\,p,f]  \; = \; \prod_{x,\mu} w^{\beta}_{j_{x,\mu}} \; ,
\label{Zdual}
\end{equation} 
where the $j_{x,\mu}$ in the link weight factors $w^{\beta}_{j_{x,\mu}}$ are computed using (\ref{hodgedecomp}). 
We remark that in $Z$ a trivial overall factor was omitted, which results from the fact that the parameterization with the 
plaquette occupation numbers is not one-to-one, since, e.g., all plaquette occupation numbers in (\ref{hodgedecomp})
can be shifted by a constant mod $q$ and the configuration of fluxes $j_{x,\mu}$ remains unchanged. 
In the dual form (\ref{Zdual}) the partition function is a sum over configurations of the plaquette variables $p_{x,\mu\nu}$ and the 
parameters $f_{\mu}$ for the winding flux. These degrees of freedom no longer need to obey any constraints. We will 
see below, that the form (\ref{Zdual}) can be used for a local Monte Carlo update of the systems, while the worm
update will be implemented directly in the worldline form (\ref{Z_WL}).

\subsection{Self-duality in two dimensions}

As a small corollary of the worldline/dual form of the $q$-state Potts model we have presented here, we re-derive 
the self-duality of the model in two dimensions \cite{pottsreview}. We work directly in the infinite volume limit,
where the contributions of the defect fluxes can be neglected. Furthermore, in 2-d we have only one type of
plaquettes, $p_{x,12}$, and we can label\footnote{Actually one could also switch to the dual lattice 
and identify the plaquettes in 2-d with the 
sites of the dual lattice. However, for our brief discussion a full implementation of the machinery of duality  \cite{savit}
is not really necessary.} these plaquettes by their lower left corner $x$, i.e., in 2-d we may denote the plaquette variables as $p_x \in \{0,1, \, ... \; q-1\}$.

The flux on the links is evaluated according to (\ref{hodgedecomp}), which in the absence of defect fluxes and in 
2-d reduces to 
\begin{equation}
j_{x,1} \; = \; ( \, p_{x} \, - \, p_{x-\hat 2} \,)
\; \mbox{mod} \; q \; \; , \quad
j_{x,2} \; = \; (\, p_{x-\hat 1} \, - \, p_x \, ) \; 
\mbox{mod} \; q \; .
\end{equation}
The values $j_{x,\mu}$ determine which of the factors
$w^{\, \beta}_{j_{x,\mu}}$ given in Eq.~(\ref{weightsw})
is taken into account. Note that 
(\ref{weightsw}) distinguishes only two values 
$w_0^{\, \beta}$ and $w_1^{\, \beta} = w_2^{\, \beta} 
= \, ... \, = w_{q-1}^{\, \beta}$. Consequently the 
weight is $w_0^{\, \beta}$ if $p_x$ and 
$p_{x-\hat \mu}$ are equal and 
$w_1^{\, \beta}$ if $p_x$ and $p_{x-\hat \mu}$ are
different. Thus we may write the 
corresponding interaction term in the form
\begin{equation}
w_1^{\, \beta} \; \left( 
\frac{w_0^{\,\beta}}{w_1^{\,\beta}}\right)^{\delta(p_x, p_{x-\hat \mu})} \; = \; 
w_1^{\, \beta} \; \, e^{\; \tilde{\beta} \;  
\delta(p_x, p_{x-\hat \mu})} \; ,
\end{equation}
where we have defined
\begin{equation}
\frac{w_0^{\, \beta}}{w_1^{\, \beta}} \; \equiv \; 
e^{\; \tilde{\beta}} \; = \;
1 \; + \; \frac{q}{e^{\,\beta} - 1} \; .
\label{betadual}
\end{equation}
Thus we may write the 2-d partition sum also as a sum
over configurations of the plaquette variables 
(an overall irrelevant constant has been dropped)
\begin{equation}
Z \; = \; \sum_{\{ p \}} \; e^{ \; \tilde\beta \sum_{x}
\sum_{\mu =1}^d \delta(p_x,p_{x+\hat\mu})}\quad, \qquad
\sum_{\{p\}} \; = \; \prod_x \, 
\frac{1}{q}\sum_{p_x=0}^{q-1} \; .
\label{Zselfdual}
\end{equation}
Comparing this form of the partition sum in terms of 
plaquette variables with the 
defining form (\ref{Zdef}) in terms of spins, we find 
that the partition function is self-dual. The
original coupling $\beta$ and the dual coupling 
$\tilde \beta$ are related via the relation 
(\ref{betadual}). 

If one now assumes that there is only a single critical 
point at $\beta_c$, then we may set 
$\tilde \beta_c = \beta_c$ and use (\ref{betadual}) 
to determine $\beta_c$, 
\begin{equation}
\beta_c \; = \; \ln\, (1+\sqrt{q}) \; .	
\end{equation}
Thus we have re-derived the well-known result \cite{betacrit} for the 
critical coupling $\beta_c$ at all values of $q$ as a 
simple corollary of the worldline representation 
presented here.

\section{Monte Carlo simulation with worldlines}

In this section we report some first exploratory studies of using the worldline formulation for Monte Carlo simulations. 
For d = 2 dimensions we discuss the implementation of the worm algorithm for the worldlines of the $q$-state Potts model and check its
correctness by comparing its results to simulations in the standard spin representation and to results of a local update 
of the dual representation. Subsequently we consider the $q = 2$ and $q = 4$ models, that both have second
order transitions, and provide a first estimate of the corresponding dynamical critical exponents 
for the worm algorithm. Finally 
we briefly discuss properties of the worm algorithm also for the first order case ($q > 4$). 

\subsection{Worm algorithm for the 
$q$-state Potts model worldline representation}

The worm algorithm for our worldline representation of the Potts model is a generalization of the Prokofev-Svistunov 
worm algorithm \cite{worm}. It starts at some randomly selected site $x_0$ of the lattice and then propagates through 
the lattice attempting to change the flux on each link it visits by $\pm 1$. The worm finishes when it reaches the 
starting point $x_0$ and all constraints are intact again. 
Each step of the worm is accepted with a Metropolis decision 
\cite{metropolis}. 

In pseudocode notation the steps of the worm can be formulated as follows ({\tt rand()} denotes a uniformly distributed 
random number in the range $[0,1]$) :

\begin{itemize}
  
\item {\tt Randomly select an increment} $\Delta \in \{-1,+1\}$

\item {\tt Randomly select a starting site} $x_0$ {\tt and set} $x \leftarrow x_0$

\item {\tt Repeat until worm terminates:}

\begin{itemize}

\item {\tt Select a direction} $\mu \in \{ \pm 1, \pm 2, \, ... \, \pm d\}$

\item {\tt Proposal flux:} $\tilde{j}_{x,\mu}  =  ( j_{x,\mu} + \, \mbox{sign} (\mu) \Delta) \; \mbox{mod} \; q\quad$ 
{\tt (with $j_{x,\mu} \equiv j_{x-\hat{|\mu|},|\mu|}$ for $\mu < 0$)}

\item {\tt Compute $\rho \; = \; w_{\tilde{j}_{x,\mu}}^{\, \beta} \, / \, w_{{j}_{x,\mu}}^{\, \beta}$}.
{\tt If (rand() $< \rho\; $)} {\tt then}  $\; {j}_{x,\mu} \leftarrow \tilde{j}_{x,\mu} \; $ {\tt and} $\; x \leftarrow  x + \hat{\mu}$

\item {\tt If ($x = x_0$)  terminate worm}

\end{itemize}

\end{itemize}

It is straightforward to see that the worm generates only admissible flux configurations and is ergodic. In order to study the 
case of non-vanishing magnetic terms one may start the worm with the insertion of a magnetic term (source) and then 
in each step of the worm also offer to close the worm with a second insertion of a magnetic term (sink). This is the version
we test below, but remark that also a variant where the worm inserts two magnetic terms connected by a 
''large hop'' is an interesting alternative option, in particular if the magnetic field is very small  
(see, e.g., the discussion \cite{Mercado:2012yf} for the special case of $q = 3$). 
Finally we remark that also other variants of the worm, such as directed worms, geometric worms or worms with 
heat-bath steps \cite{worm_suwa,worm_hitchcock,giuliani} could further 
extend the toolbox for worldline simulations of $q$-state Potts models. 

 \begin{figure}[p]
\begin{center}
\hspace*{0mm}
\includegraphics[width=160mm,clip]{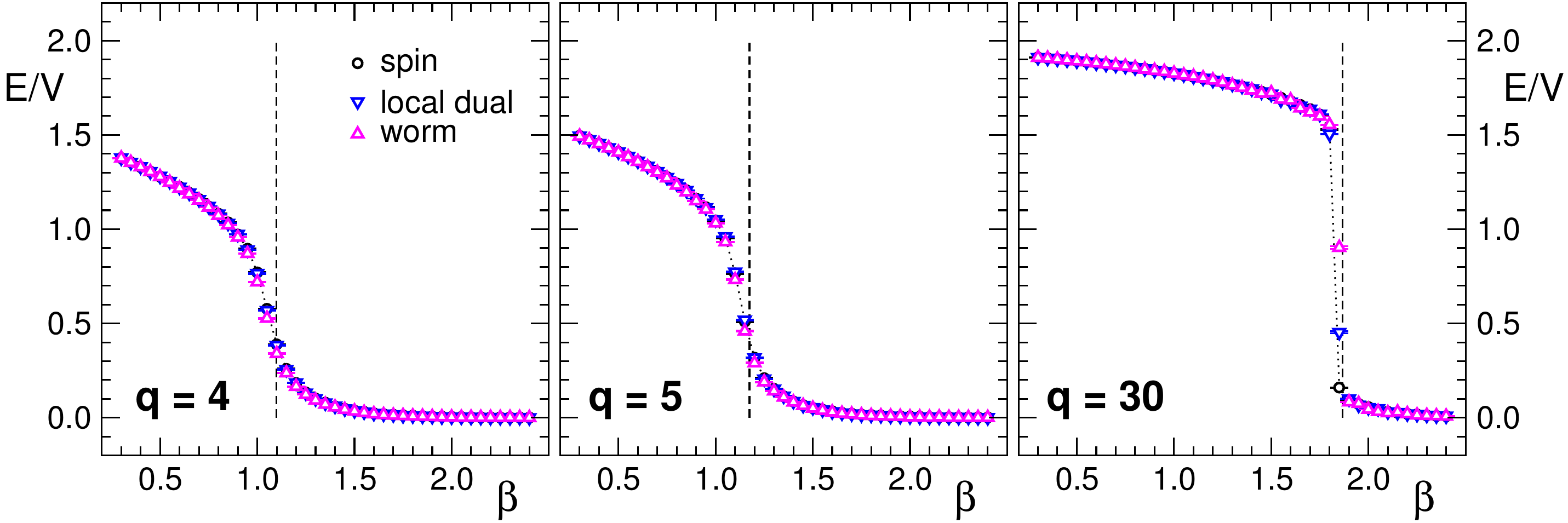}
\vskip6mm
\includegraphics[width=160mm,clip]{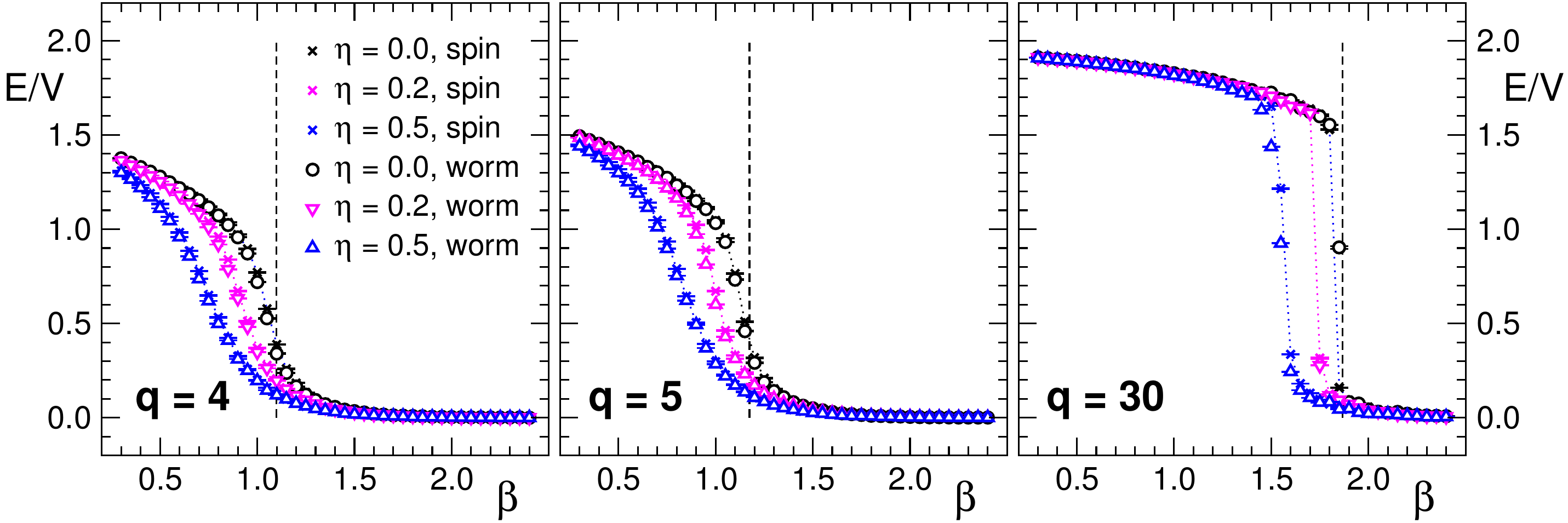}
\end{center}
\caption{Top: Comparison of Monte Carlo results for the energy density as a function of $\beta$ for $q = 4$, $q = 5$ and 
$q = 30$ (left to right). We show data for a simulation in the spin representation (circles), a local dual simulation 
(blue triangles pointing down), as well as data from the worm simulation (magenta triangles pointing up). 
The dashed vertical lines mark $\beta_c$.
We use small $8 \times 8$ lattices where finite volume effects are still prominent in order to test if the 
dual update properly populates the defect lines and if the worm indeed produces the necessary 
winding worldlines.  
Bottom: Results for the energy density at finite external field for the $q=4$, $q = 5$ and $q=30$ models (left to right). 
We compare the data from the local spin simulation (crosses) to those from the worm algorithm (other symbols). 
Again the lattice size is  $8 \times 8$ for assessing finite volume effects and we show results for $\eta = 0.0$,
$\eta = 0.2$ and $\eta = 0.5$. 
\label{comp_fig}}	
\end{figure}  

Before we come to a more detailed assessment of Monte Carlo simulations of the worldline representation of the 
$q$-state Potts model we briefly discuss the evaluation of the new representation and the worm algorithm. 
In the top row of plots of Fig.~\ref{comp_fig} we compare the results for the energy density as a function of $\beta$ 
from three different simulations based on the conventional spin representation (circles), the local dual 
representation (\ref{Zdual}) (blue triangles pointing down) and the worm algorithm discussed in this section 
which updates the worldline representation (magenta triangles pointing up). We compare the models
for $q=4$, $q=5$ and $q = 30$ (left to right) to assess second order, weak first-order and strong first order behavior.
We choose a small lattice of size $8 \times 8$ where finite volume effects are still strong. This allows us to test also 
the non-trivial topological aspects that play a role in the dual representation via the defect fluxes, and in the worldline
representation through worms that wind around the periodic boundary conditions. 
Obviously the results of all three simulations agree very well for the whole range of $\beta$ and the different $q$ we 
consider.   

A similar picture results for the comparison of local spin and worm simulations for the case of finite magnetic field shown 
in the bottom row of plots in Fig.~\ref{comp_fig}. Here the worm was started and terminated with magnetic flux insertions
as discussed above. Again we compare the energy density as a function of $\beta$ for $q = 4$, $q = 5$ and 
$q = 30$ (left to right) on a small $8 \times 8$ lattice to assess finite volume effects. We use $\eta = 0.0$, $\eta = 0.2$ 
and $\eta = 0.5$ and again find very good agreement of local spin simulation results (crosses) with those from the 
worm algorithm (other symbols).  

\subsection{Exploratory assessment of the algorithm for
$q \leq 4$ (2$^{nd}$ order transitions)}

We start the analysis of the worm algorithm with exploratory studies in the $q = 2$ and $q = 4$ cases where the 
Potts model has second order phase transitions at the critical couplings $\beta_c = \ln(1 + \sqrt{q})$. The key challenge
for Monte Carlo simulations is critical slowing down near $\beta_c$ that can be quantified by the dynamical critical 
exponent $z$. With finite volume scaling techniques for simulations done at $\beta_c$ we here estimate $z$ for the 
worm algorithm in the $q = 2$ and $q=4$ cases using the energy 
as observable. Typical statistics for these determinations 
are $10^6$  to $10^{11}$ configurations, which reflects the demanding nature of the determination of dynamical
critical exponents. 

For completeness we summarize the conventions of autocorrelation functions, autocorrelation times and the dynamical 
critical exponent $z$. We remark that for a full assessment of the autocorrelation times a complete set of observables 
should be considered. In our exploratory study we here restrict ourselves to the energy as our main observable and stress 
that other observables could have longer autocorrelation times. The normalized autocorrelation function for the energy is
\begin{equation}
	A_E(k) \; = \; \frac{\langle E_i \ E_{i+k} \rangle - 
	\langle E_i\rangle \langle E_i\rangle}{\langle E_{i}^{2}\rangle - \langle E_i\rangle \langle E_i\rangle} \ ,
\end{equation}
where $E_i$ is the value of the energy at the $i$-th step of the Monte Carlo time series.  
The autocorrelation function decays exponentially for sufficiently large time separations $k$ and we may use this
exponential decay to define the exponential autocorrelation time $\tau^{(E)}_{exp}$. Summing the autocorrelation function 
defines the integrated autocorrelation time $\tau^{(E)}_{int}$, such that the defining equations for the two 
autocorrelation times we use here are given by \cite{sokal},
\begin{equation}
	A_E(k) \; \propto \; \exp \left( {\frac{-k}{\tau^{(E)}_{exp}}}\right) \; \; \mbox{for} \; \; k \, \rightarrow \, \infty \; \; ,
	\qquad 
	\tau^{(E)}_{int} \; = \; \frac{1}{2} + \sum_{k=1}^{k_{max}} A_E(k) \; ,
	\label{equ:tau}
\end{equation}
where $\tau^{(E)}_{int} \leq \tau^{(E)}_{exp}$. Note that in principle $k_{max}$ should be considered in the limit 
$k_{max} \rightarrow \infty$, but in a practical determination one needs to cut off the sum that defines 
$\tau^{(E)}_{int}$. Such a cut-off is considered safe when the summation obeys 
$k_{max} \geq 6 \, \tau^{(E)}_{int}\, (k_{max})$ \cite{sokal,janke}. 

In the critical region of second order transitions the autocorrelation time diverges as a power of the correlation length $\xi$, 
\begin{equation}
	\tau \; \sim \; \xi^z \; \sim \; L^z \ ,  
	\label{equ:dynamic_exponent}
\end{equation} 
where $z$ is the dynamical critical exponent, that can be different for different observables (and different algorithms of course). 
On a finite lattice of volume $V = L^d$ (here $d = 2$), the correlation length 
$\xi$ cannot diverge and will instead be cut off by the linear extent of the lattice $L$, which is expressed in the second 
equality\footnote{We remark that there are also sub-leading terms that correct the leading behavior 
(\ref{equ:dynamic_exponent}) such that they could influence the results. Thus in principle a full error analysis would have to study 
stability of the fit results when including such terms. This is a step omitted in our exploratory study here.} of (\ref{equ:dynamic_exponent}).

\begin{figure}[t] 
	\centering
	\includegraphics[width=130mm,clip]{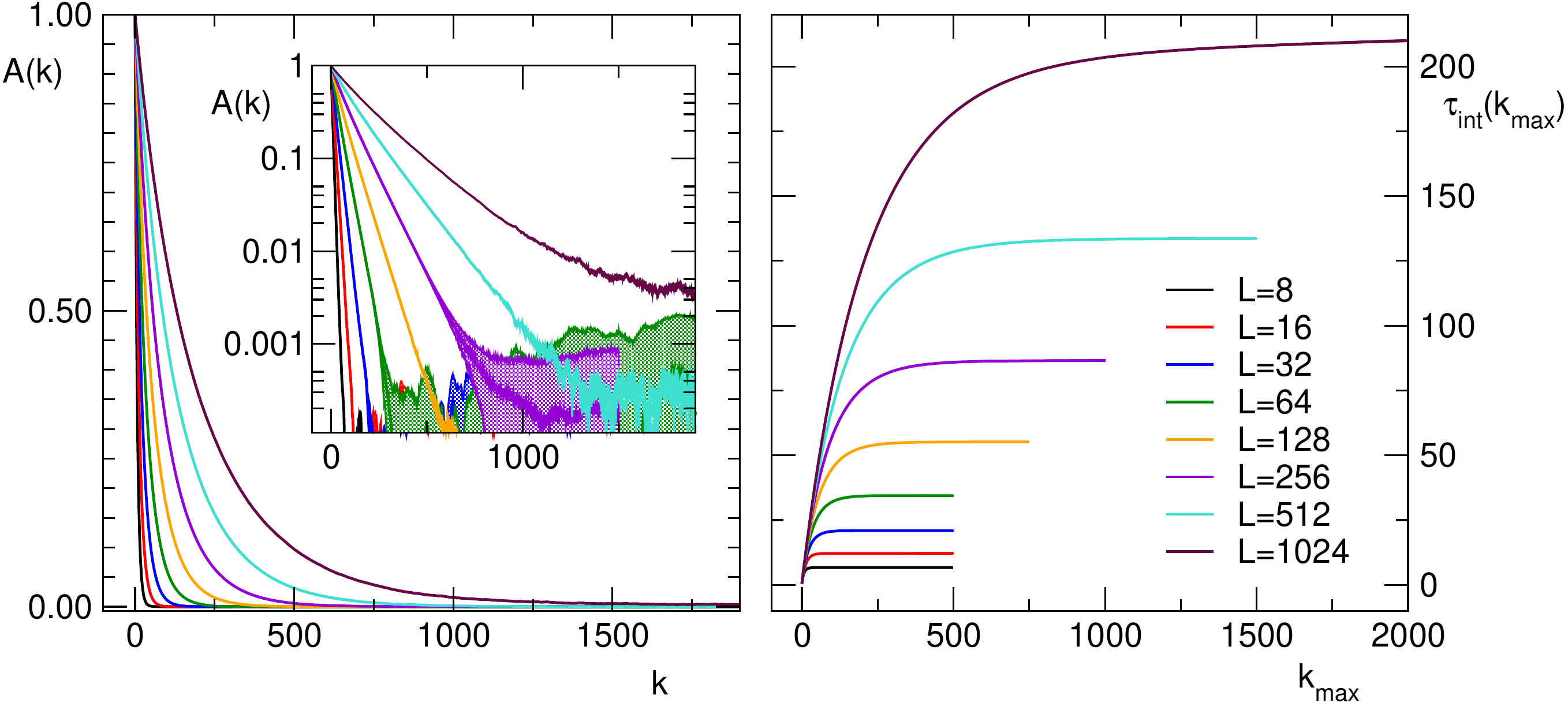} 
	\vskip3mm
	\includegraphics[width=130mm,clip]{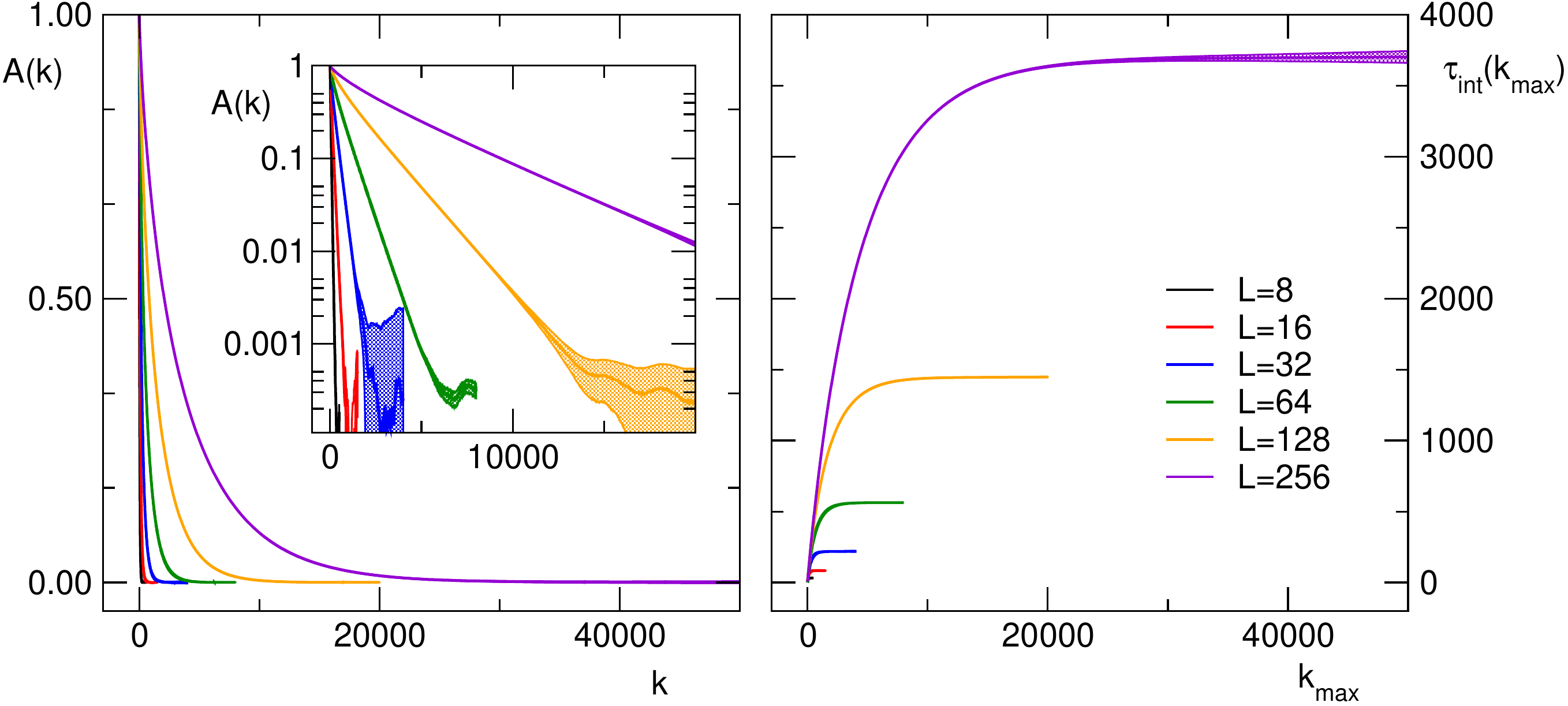} 
	\caption{Autocorrelation analysis for $q=2$ (top row of plots) and $q = 4$ (bottom). 
	Lhs.: Plot of the normalized autocorrelation function for different volumes $L^2$ as a function 
	of the Monte Carlo time separation $k$ (logarithmic scale on the vertical axis for the inserted plots). 
	Rhs.: Plot of the integrated autocorrelation time as a function of the cut-off 
	$k_{max}$, again for different volumes $L^2$ and $q=2$ (top) as well as $q = 4$ (bottom).}
	\label{fig:auto_analysis}
\end{figure}

For comparing autocorrelation times to the Metropolis algorithm or the Swendsen-Wang cluster algorithm, where one 
update sweep visits all degrees of freedom of the lattice, we need to define proper units for the autocorrelation times.
 We take this into account by normalizing the measured autocorrelation times with the factor 
 $\langle \mbox{\tt worm length} \rangle / (2V)$, i.e., the average fraction of the degrees of freedom that was subject 
 to a worm update. 
 
In the lhs.\ plots of Fig.~\ref{fig:auto_analysis} we show results for the normalized autocorrelation function 
$A_E(k)$ as a function of the Monte Carlo time separation $k$. We compare the autocorrelation functions for different 
volumes $L^2$ and consider $q=2$ (top) and $q=4$ (bottom). The inserted plots show the same data using a 
logarithmic scale for the vertical axis. We observe exponential behavior of the correlation function up to 
values of $k$ where the autocorrelation function becomes smaller than the statistical error from the Monte
Carlo simulation. 

From the results for the autocorrelation function shown in Fig.~\ref{fig:auto_analysis} we now can determine 
$\tau^{(E)}_{int}$ and $\tau^{(E)}_{exp}$ according to (\ref{equ:tau}). The rhs.\ plots of Fig.~\ref{fig:auto_analysis} illustrate
our determination of $\tau^{(E)}_{int}$: We show the integrated autocorrelation time $\tau^{(E)}_{int}(k_{max})$ 
as a function of the cut-off $k_{max}$ and again compare $q = 2$ (top) and $q = 4$ (bottom). Obviously, when 
considered as a function $\tau^{(E)}_{int}(k_{max})$ of the cut-off $k_{max}$, the integrated autocorrelation time 
approaches a constant for large $k_{max}$, which is an estimate for $\tau^{(E)}_{int}$. 
Thus we estimate $\tau^{(E)}_{int}$ by using the large-$k_{max}$ values of $\tau^{(E)}_{int}(k_{max})$ 
self-consistently for all volumes with the condition $k_{max} \geq 6 \tau^{(E)}_{int}(k_{max})$ and denote the 
results as $\tau^{(E)}_{int,cut}$. 

We explore a second strategy for determining $\tau^{(E)}_{int}$ by using a fit of $\tau^{(E)}_{int}(k_{max})$ in the low-$k_{max}$
range up to where $\tau^{(E)}_{int}(k_{max})$ starts to saturate to the form \cite{janke}
$\tau^{(E)}_{int}(k_{max}) = \tau^{(E)}_{int} - c \exp(-k_{max}/\tau^{(E)}_{exp})$. The fit parameters are $\tau^{(E)}_{int}$ at $k_{\max} = \infty$,
the exponential autocorrelation time $\tau^{(E)}_{exp}$ and an irrelevant constant $c$. This leads to estimates 
of both, the integrated and the exponential autocorrelation time which we denote by $\tau^{(E)}_{int,fit}$ and 
$\tau^{(E)}_{exp,fit}$, respectively.

\begin{figure}[t] 
	\centering
	\includegraphics[width=130mm,clip]{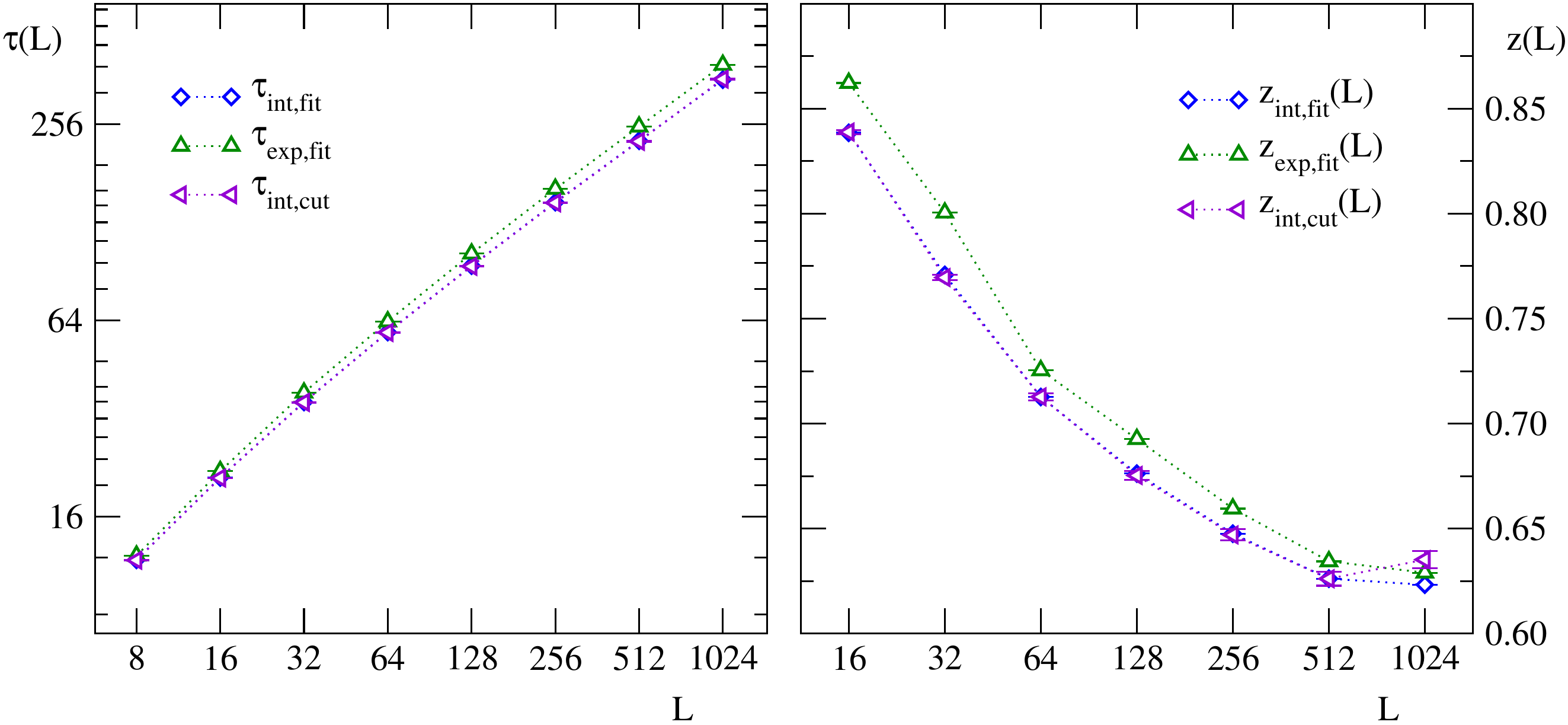}
	\vskip3mm
	\includegraphics[width=130mm,clip]{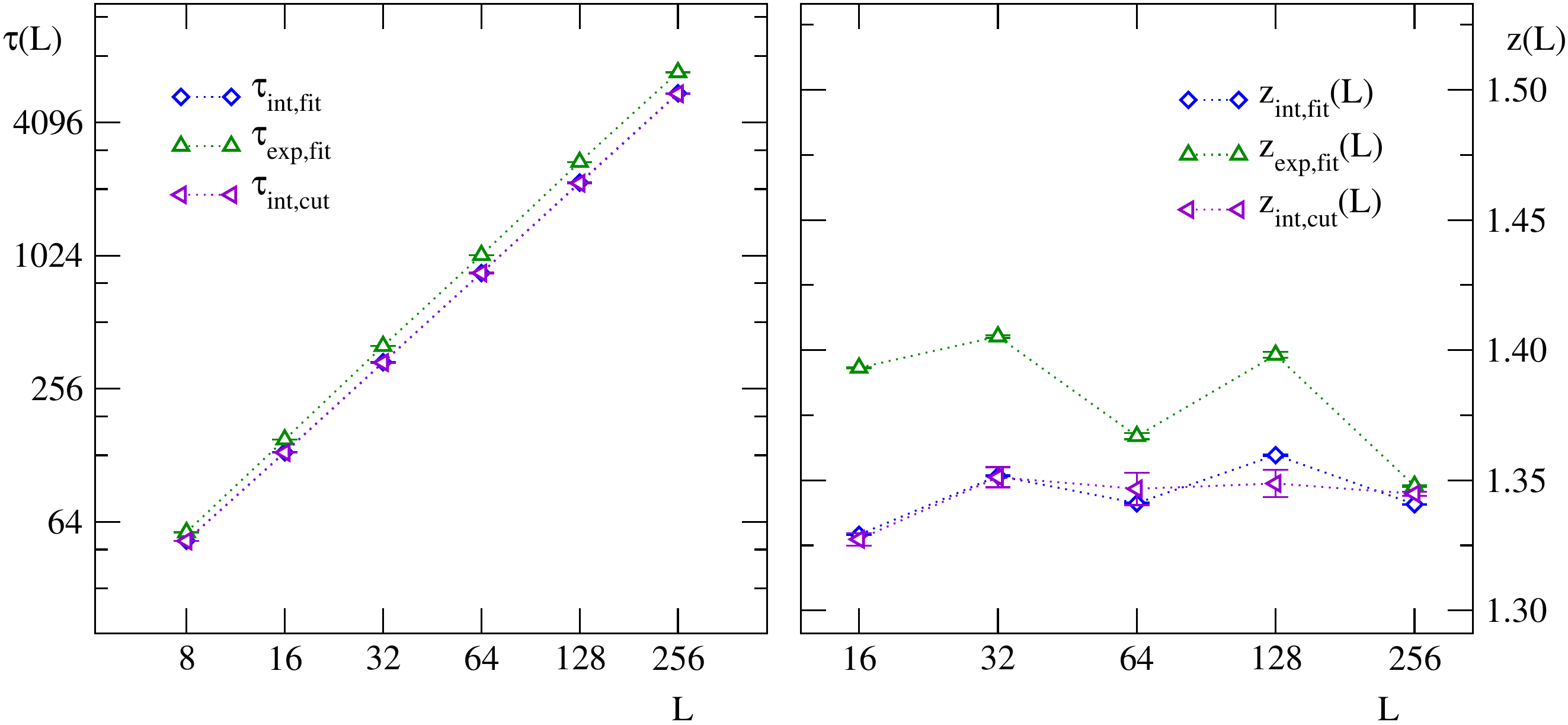}	 
	\caption{Autocorrelation times and critical exponents for $q=2$ (top row of plots) and
	$q = 4$ (bottom). Lhs.: Plot of the results for the autocorrelation times as function of the lattice extent $L$.
	Rhs.: Results for the dynamical critical exponents obtained form fits of $\tau$ according to 
	(\ref{equ:dynamic_exponent}) using pairs $L_0<L$ of subsequent lattice sizes.}
	\label{fig:z_analysis}
\end{figure}

In the lhs.~plots of Fig.~\ref{fig:z_analysis} we show our results for $\tau^{(E)}_{int,cut}$, $\tau^{(E)}_{int,fit}$ and $\tau^{(E)}_{exp,fit}$ 
as a function of the lattice extent $L$, using a double logarithmic plot (top: $q = 2$, bottom: $q = 4$). 
We observe a slight deviation from perfect linear behavior expected according to (\ref{equ:dynamic_exponent}). 
We attribute these deviations to finite size corrections, and in order to take them into account we fit the data 
for $\tau$ from consecutive pairs of volumes $V_0 = L_0^{\,2}$ and $V = L^2$ with extents 
$L_0<L$ with (\ref{equ:dynamic_exponent}). These fits give rise to results for the dynamical critical exponents 
$z$ which we study as a function of the larger extent $L$.  

On the rhs.~of Fig.~\ref{fig:z_analysis} we plot $z$ as a function of $L$, showing our results for all 
three determinations of the autocorrelation time $\tau^{(E)}_{int,cut}$, $\tau^{(E)}_{int,fit}$ and $\tau^{(E)}_{exp,fit}$. 
For the $q = 2$ case (top right) the estimates for $z$ decrease with increasing spatial extent $L$, eventually
reaching saturation which indicates that with our largest volumes we are reaching the infinite volume extrapolated result 
for $z^{(E)}$ and all three determinations consistently give rise to values of $z \sim 0.63$.  Since the 
dynamical critical exponent usually is determined from the integrated 
autocorrelation time, we here quote the values $z^{(E)}_{int,cut}=0.6352(2)$ and $z^{(E)}_{int,fit}=0.62325(2)$ which we take from the 
largest volume pair, and stress that the errors given are only the statistical errors. In order to estimate also the systematical error
we average these results into our final result 
$z^{(E)}_{int} = 0.63(1)$, where we quote the distance of the two different determinations as our systematical error (the 
statistical error is much smaller). For completeness we also quote our final result for the exponential autocorrelation time,  
$z^{(E)}_{exp} = 0.629(1)$ (error is only the statistical error).

When inspecting the $q = 4$ results for the dynamical critical exponent (bottom right of Fig.~\ref{fig:z_analysis})
plotted against $L$ on a double logarithmic scale, we do not observe a systematic variation with the spatial 
extent $L$. Quoting again the results from the largest volume pair we find $z^{(E)}_{int,cut}=1.3448(7)$ and 
$z^{(E)}_{int,fit}=1.34076(8)$ where again the errors are only the statistical errors. As before we average these to our final 
value of $z^{(E)}_{int}=1.34(1)$, where the error is an estimate of the systematic error determined as the 
difference between the two determinations. The result for the dynamical critical exponent 
determined from the exponential autocorrelation time is $z^{(E)}_{exp} = 1.348(1)$.

In summary, for the simple worm algorithm we find critical dynamical exponents $z^{(E)}_{int} = 0.63(1)$ for $q = 2$ and 
$z^{(E)}_{int} = 1.34(1)$ for $q =4$, but remark again that the systematical errors could be larger, e.g., from contributions
due to sub-leading terms in (\ref{equ:dynamic_exponent}). Obviously the simple worm studied here is still outperformed by the corresponding 
Swendsen Wang cluster algorithms with $z < 0.3$ for $q = 2$ and $z \sim 1.0$ for $q = 4$ \cite{salas}. 
This may be attributed 
to effects such as links being visited several times by the worm \cite{worm_suwa}. For improved worm techniques
\cite{worm_hitchcock,worm_suwa} dynamical critical exponents of $z < 0.3$ were reported for $q = 2$ and we expect 
that such an increase in performance with improved worms should be possible also for the
$q = 3$ and $q = 4$ models. Concerning the 
dependence on $q$ the increase from  $z^{(E)}_{int} \sim 0.63$ for $q = 2$ to $z^{(E)}_{int} \sim 1.34$ for $q =4$ 
indicates that as $q$ is increased and one approaches the first order regime, the numerical simulations 
become more demanding. We remark that a similar observation was also made for the Swendsen-Wang cluster 
algorithm \cite{salas}.

We conclude with a more general comment on the comparison of algorithms that work with two different representations, such as the spin 
and the worldline representation in this study\footnote{We thank the referee for pointing out to us the aspects discussed in this paragraph.}. 
Certainly the estimators for observables, such as the energy used in our study, 
have the same mean value in both representations. However, the squares of these 
observables will in general not coincide for the different representations (actually one can show that the heat capacity has an extra term 
in the worldline representation). This allows for different variances, which may even scale differently with the lattice size. Hence the method
with larger $z$ could in principle still yield the more accurate results. Thus the comparison of simulations in different representations can be 
expected to be more complicated than just ranking $z$, which thus should be considered only as a first assessment.

\subsection{Challenges for $q > 4$ simulations (1$^{st}$order transitions)}

For $q>4$ the 2-d Potts model has first order transitions at $\beta_c = \ln(1 + \sqrt{q})$ and for reasonably large 
volumes Monte Carlo algorithms face a severe sampling problem due to mixing of phases at $\beta_c$. This issue
typically can be addressed with various reweighting and histogram techniques 
(see, e.g., \cite{barkema,landau}), which of course can also be implemented for worm 
algorithms updating the worldline representation. 

We here do not attempt a detailed and systematic analysis of worm algorithms in the first order regime of the Potts 
models, but only briefly report some simple observations we made in small test simulations at $q > 4$, partly using large 
$q$ where the first order transition is very strong.

\begin{figure}[t] 
	\centering
	\hspace*{-10mm}
	\includegraphics[height=65mm,clip]{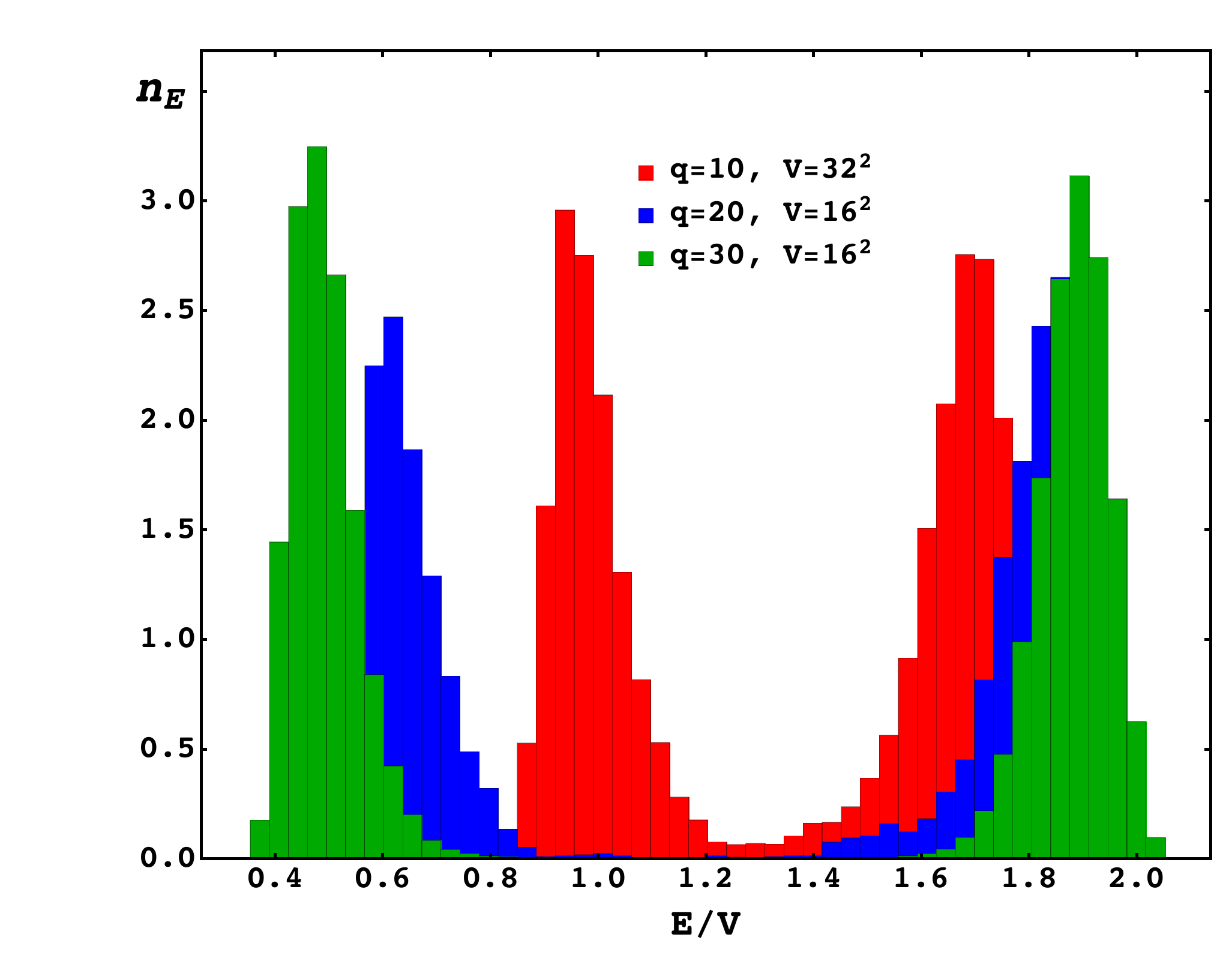} \hspace{10mm}
	\includegraphics[height=70mm, clip]{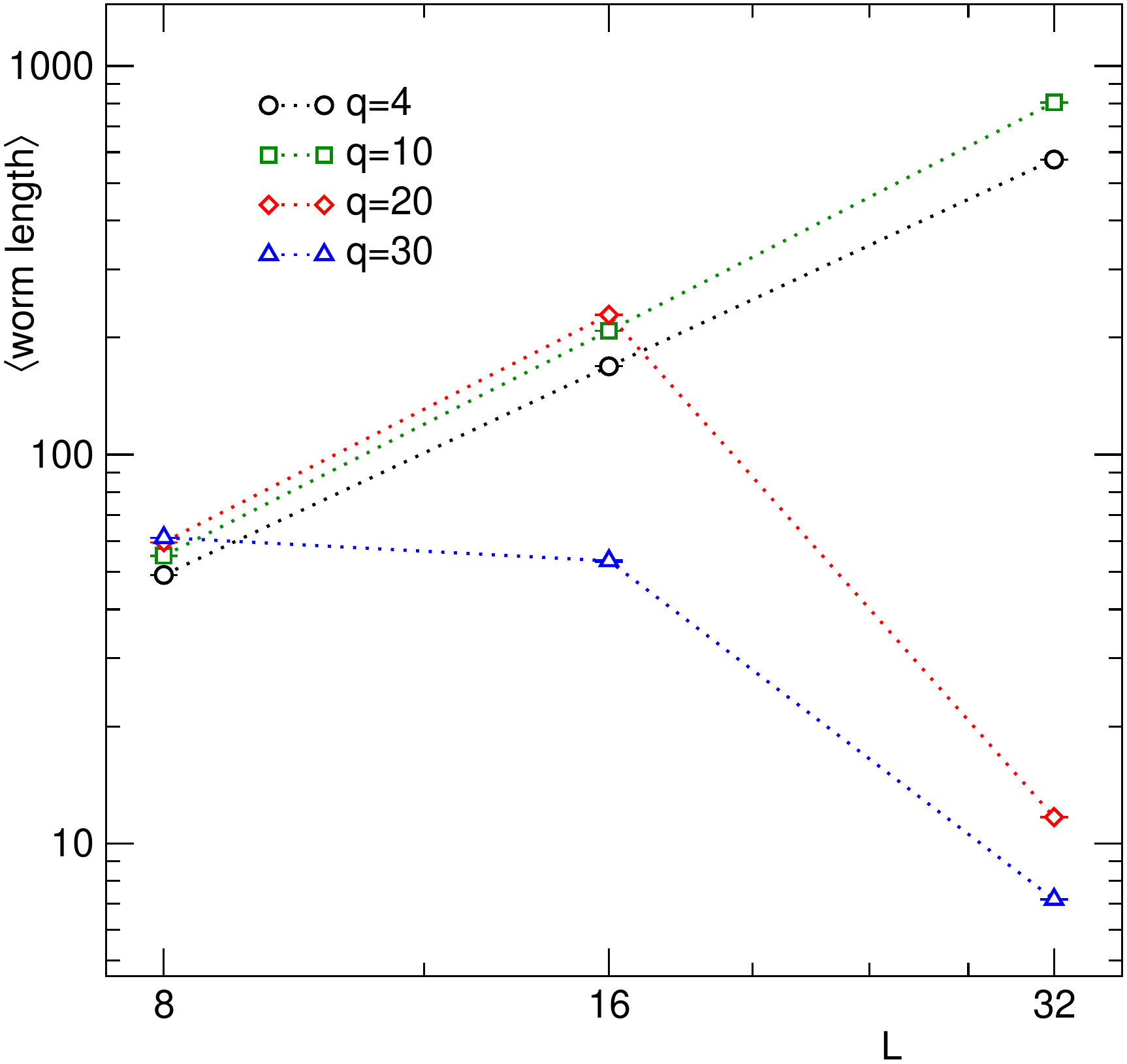} 
	\caption{Lhs.: Histograms for the energy per volume for the $q=10$, $q = 20$ and $q = 30$ models. Rhs.: 
	Average worm length as a function of the spatial extent $L$. We compare the results for $q = 4$ 
	(second order behavior) with $q=10$, $q = 20$   and $q = 30$ where the first order behavior becomes 
	increasingly stronger.}
	\label{fig:histogram}
\end{figure}

The sampling problem is illustrated in the lhs.~plot of  Fig.~\ref{fig:histogram}, where we show histograms for 
the energy density $E/V$ computed at $\beta_c$. We compare the histograms for $q=10$, $q=20$ and 
$q=30$ (note that due to the mentioned sampling problems a smaller volume was used for the latter two) 
that display the characteristic 
double peak structure of first order transitions. We observe, that as $q$ is increased 
and the transition becomes harder, the region between the peaks becomes wider, or -- in more
physical terms -- the latent heat grows. Thus it is increasingly harder for an algorithm that samples the 
un-modified Boltzmann weight to cross between the two peaks and simulations near $\beta_c$
get stuck in only a part of configuration space. 

It is interesting to see how the sampling problem affects the worm algorithm for the worldline representation. 
This is illustrated in the rhs.\ plot of Fig.~\ref{fig:histogram} where we show for simulations at $\beta_c$ the average worm length as a function of $L$ for different $q$ on a double logarithmic scale. For the second order 
case ($q = 4$) the worm length keeps
growing with $L$ (at least up to the volumes we considered), while for $q = 20$ and $q = 30$ it starts to drop for
the larger values of $L$ which we attribute to the fact that between the peaks the distribution in the histogram becomes 
suppressed exponentially and obviously the worm becomes confined in only one of the peaks. For $q = 10$ we
observe a behavior that is similar to the $q = 4$ case, probably because at $q = 10$ the first order transition is not so strong yet. 
However, one may expect that also for $q = 10$ the worm length will start to drop for sufficiently large $L$ and we conclude 
that the well-known sampling problems near first order transitions manifest themselves in a considerable decrease of 
the average worm length which reflects insufficient sampling. 

\section{Summary and discussion}
In this paper we have explored a worldline representation for the $q$-state Potts model with magnetic term in
arbitrary dimensions. The worldlines are described by link-based flux variables and the flux 
is conserved modulo $q$. For non-zero magnetic
field the worldlines can start and end in magnetic source and sink terms. These are absent for vanishing
magnetic field and admissible configurations are closed loops of flux that is conserved modulo $q$. For this
case we show that one may resolve the constraints by 
introducing dual variables that consist of flux around plaquettes and defect fluxes
that wind around the periodic boundaries, i.e., we implement the Hodge decomposition. 
Finally we show that in the 2-d case one may use the dual representation
to establish self-duality of the $q$-state Potts models as a simple corollary and re-derive the known critical 
couplings $\beta_c = \ln(1+\sqrt{q})$. 

Having established the worldline and the dual representation we address possible Monte Carlo simulations in 
the new representations. We present a worm algorithm for updating the worldlines and verify it against a 
conventional simulation in terms of spins and a local simulation of the dual representation. These tests were
done for both, vanishing and finite external field and
not only check the correct implementation of the worm algorithm, but also the correctness of the worldline and the dual 
representations. In an exploratory study we determine the dynamical critical exponent of the worm algorithm
in two dimensions for the $q=2$ and $q=4$ cases. The corresponding results are $z = 0.629(6)$ and 
$z = 1.342(2)$. 
A brief look at the models with $q > 4$ indicates that the sampling problem near the corresponding first order 
transitions leads to a decrease of the worm length at larger volumes. This suggests that also the worm becomes
trapped in only a part of the configuration space and suitable reweighting and histogram techniques need to
be used.

It is obvious that the numerical assessment of the worldline simulations in this paper has only an exploratory
character and several future directions will be interesting to follow. In particular using more advanced worm 
techniques for studies of the $q > 2$ cases in dimensions larger than d = 2, as well as extended studies of the system 
with non-vanishing magnetic field terms are planned for future work.

\vskip5mm
\noindent
{\bf Acknowledgments:} 
Parts of the numerical analysis were done at the VSC clusters in Vienna, and we are grateful for support and 
computer time at VSC. This work is partly supported by the Austrian Science 
Fund FWF Grant.\ Nr.\ I 2886-N27.

\end{document}